\documentclass[aps,prb,twocolumn]{revtex4} %,preprint %revtex4-2

\usepackage{graphicx}% Include figure files
\usepackage{dcolumn}% Align table columns on decimal point
\usepackage{bm}% bold math
\usepackage{amsmath} % subequations \begin{subequations}...\end{subequations}

\newcommand{\comment}[1]{}
\begin{document}
%\preprint{}   % Preprint number in upper right corner
%\renewcommand{\theequation}{\arabic{section}.\arabic{equation}}

\title{Thermodynamic Explanation of the Meissner-Ochsenfeld Effect
in Superconductors}

%\author{}
%\thanks{}

\author{Phil Attard}
\affiliation{ {\tt phil.attard1@gmail.com}  19--26 Aug 2025}
%\\ 23 April 2025}
%\noindent {\tt  Projects/QSM25/Meissner/Meissner.tex}
%\affiliation{\protect\texttt{phil.attard1@gmail.com}}

%\date{\today. Begun  17 July, 6 Aug, 19 Aug 2025
%phil.attard1@gmail.com
%\\
%notes in Projects/QSM25/May25.tex:\S~6G (31Jul25), Aug25.tex}
% no notes for third version, 19 Aug 25

\begin{abstract}
The thermodynamic principle of superfluid flow
---that the energy is minimized at constant entropy---
%chemical potential is uniform in space
is applied to superconducting currents
to derive the Meissner-Ochsenfeld effect
in which magnetic fields are expelled from superconductors.
The principle gives a modified form for the first London equation
that does not trap magnetic fields within a superconductor.
The physical mechanism by which a critical magnetic field
destroys superconductivity is identified.
\end{abstract}

%\pacs{}
%\keywords{}

\maketitle

%\newpage
%%%%%%%%%%%%%%%%%%%%%%%%%%%%%%%%%%%%%%%%%%%%%%%%%%%%%%%%%%%%%%%%%%%%%%%%%%
%
%\section{Introduction}
\setcounter{equation}{0} \setcounter{subsubsection}{0}
%%\renewcommand{\theequation}{\arabic{equation}}
%\renewcommand{\theequation}{\arabic{section}.\arabic{equation}}
%\renewcommand{\theequation}{\Alph{section}.\arabic{equation}}
%
%%%%%%%%%%%%%%%%%%%%%%%%%%%%%%%%%%%%%%%%%%%%%%%%%%%%%%%%%%%%%%%%%%%%%%%%%%

%%%%%%%%%%%%%%%%%%%%%%%%%%%%%%%%%%%%%%%%%%%%%%%%%%%%%%%%%%%%%%%%%%%%%%
%\subsection{Two-Fluid Theory}

This paper aims to derive and to explain the Meissner-Ochsenfeld (1933)
effect wherein a superconductor expels
a magnetic field from its interior
up until a critical field that destroys the superconductivity
(Annett 2004, Kittel 1976, Tinkham 2004).
As the latter limits the power of superconducting electromagnets,
which is one of the main applications of superconductivity,
there is some motivation to understand the cause of the effect.
%at the molecular-level,
%has the potential to guide the design of better superconducting materials.

Although the Meissner-Ochsenfeld effect
has been measured and quantified in great detail
for a variety of superconductors,
there appears to be nothing known about its physical basis
(Annett 2004, Kittel 1976, Tinkham 2004).
Since it takes work by the system to magnetize the superconductor
so as to cancel the applied field in its interior,
it is reasonable to ask what compensation is gained
by the system in doing so.
That the free energy stabilizing the superconducting state compared
to the normal state can be related to the critical applied field
(Annett 2004, Kittel 1976, Tinkham 2004)
does not explain
why the superconducting state is incompatible with any magnetic field.

Equally unsatisfactory as an explanation
is the second London equation for superconductivity
(F and H London 1935),
which is an empirical equation that accounts for the absence
of any nett magnetic field in the interior of the superconductor,
and for the rate of decay of such fields in the surface region.
It was designed to be consistent with the Meissner-Ochsenfeld effect
rather than to explain it.
%and it does not predict the existence of a critical applied field.
The second London equation has been taken as the axiomatic
basis of superconducting currents
without offering an underlying physical mechanism for the equation
or for the Meissner-Ochsenfeld effect.
The results in this paper suggest that the London equations
are a consequence of the thermodynamics of superconducting systems
rather than the fundamental axiomatic basis for their behavior.

Superconductivity and superfluidity
are related by Bose-Einstein condensation
(F London 1938).
Cooper (1956) showed that pairs of electrons with opposite spin
bound by an attractive potential form an effective boson.
In the quantum mechanical BCS theory
(Bardeen Cooper and Schrieffer 1957),
the  weak and long-ranged binding
arises from coupling with lattice vibrations,
which is the operative mechanism
for low-temperature superconductivity
(Annett 2004, Tinkham 2004).
For high-temperature superconductivity,
the present author believes that
quantum statistical mechanics prevails,
and that the paired electrons are bound
by the primary minimum at short range
in the oscillatory potential of mean force
that occurs at high coupling in charged fluids
(Attard 2022b, 2025a Ch.~6).
Other mechanisms for high-temperature superconductivity,
akin to the quantum mechanical BCS theory,
have been proposed
(see references in Annett (2004), Attard (2022b, 2025a Ch.~6),
and Tinkham (2004)).
The other difference between the BCS theory and the present author's approach
is that Cooper pairs have no momentum (Cooper 1956),
whereas the author's bosonic electron pairs have nett momentum
(Attard 2022b, 2025a Ch.~6).
The present derivation of the Meissner-Ochsenfeld effect %(1933)
does not depend upon the precise binding mechanism or momentum
of an electron pair.

%%%%%%%%%%%%%%%%%%%%%%%%%%%%%%%%%%%%%%%%%%%%%%
\subsubsection{Meissner-Ochsenfeld Effect}

The energy of an electron in a magnetic field depends upon its spin,
\begin{eqnarray}
\varepsilon_\pm
& = & \pm  \mu_{\rm B} B
\nonumber \\ & = &
 \pm  \mu_{\rm B}( B_{\rm ap} + \mu_0 M/V)
\nonumber \\ & = &
 \pm  \mu_{\rm B} (1+\chi) B_{\rm ap} .
\end{eqnarray}
Here in SI units
$\mu_0$ is the permeability of free space,
$\mu_{\rm B} = e \hbar /2m$ is the Bohr magneton,
$e$ is the charge on a proton,
$\hbar = h/2\pi$ gives Planck's constant,
and $m$ is the mass of an electron.
Also ${\bf B}_{\rm ap}$ is the applied magnetic field,
${\bf B}$ is the local magnetic field,
and ${\bf M}/V$ is the induced magnetization per unit volume
(Kittel 1976, Pathria 1972).
The quantity $\chi$ is the magnetic susceptibility per unit volume
of the medium.

In the present case the magnetic susceptibility
is due to induced superconducting currents
rather than to Pauli paramagnetism, for example,
which is relatively negligible.
The goal of the present analysis is to show
that superconductors must be perfectly diamagnetic,
$ \chi = -1 $.

%Also ${\bf a}={\bf n} \Delta_p$ is the quantized momentum,
%with ${\bf n}$ being a three-dimensional vector of integers,
%and $\Delta_p = 2\pi \hbar/L$ being the spacing between momentum states,
%with the volume of the cube that is the system being $V=L^3$.
%We take the lower energy state to correspond to spin-up.

Because the electrons have equal and opposite spin,
a bosonic electron pair with separation
${\bf q}_2={\bf q}_+-{\bf q}_-$
is a magnetic quadrupole.
In a local magnetic field ${\bf B}({\bf r})$
the magnetic contribution to the pair energy is
\begin{eqnarray}
\varepsilon_2({\bf r})
& = &
\mu_{\rm B} {\bf q}_2 \cdot \nabla {B}({\bf r})
\nonumber \\ & = &
\frac{-\beta \mu_{\rm B}^2 \overline{q}_2^2}{3}(\nabla B({\bf r}))^2 .
\end{eqnarray}
The second equality follows after a classical average
and linearization for weak fields,
with the inverse temperature being $\beta = 1/k_{\rm B}T$.

Below we shall include as well an electric field.
The present analysis is for a macroscopic region of the superconductor,
and we suppose that any potential difference across such a region
would be short-circuited by rearrangement of the superconducting electrons.

\comment{ %%%%%%%%%%%%%%%%%%%%%%%%%%%%%%%%%%%%%%%%%%%%%%%%%%%%%%%%
If the gradient is parallel to the $z$-axis,
and if $\alpha \equiv \beta \mu_{\rm B} \overline{q}_2 B'$,
then
\begin{eqnarray}
\langle {\bf q}_2 \rangle
& = &
\frac{
\hat{\bf z}\int_{-1}^1 {\rm d}x\;
e^{-\alpha x } \overline q_2 x
}{
\int_{-1}^1 {\rm d}x\;
e^{-\alpha x }
}
\nonumber \\ & = &
\hat{\bf z} \overline{q}_2
\frac{
2 \cosh \alpha
-
\int_{-1}^1 {\rm d}x\;
e^{-\alpha x }
}{
-2 \sinh \alpha }
\nonumber \\ & = &
\hat{\bf z} \overline{q}_2
\frac{
2 \cosh \alpha - 2\alpha^{-1}\sinh \alpha
}{
-2 \sinh \alpha }
\nonumber \\ & \approx &
\hat{\bf z} \overline{q}_2
\frac{
1 + \alpha^2 /2  -  [ 1 + \alpha^2/6 ]
}{
-\alpha }
\nonumber \\ & = &
\frac{-\overline{q}_2\alpha}{3}\hat{\bf z}
\nonumber \\ & = &
\frac{-\beta \mu_{\rm B} \overline{q}_2^2 B'}{3}\hat{\bf z} .
\end{eqnarray}
Hence
\begin{eqnarray}
\varepsilon_2({\bf r})
& \approx &
\frac{-\beta \mu_{\rm B}^2 \overline{q}_2^2}{3}(\nabla B)^2 .
\end{eqnarray}
} % end comment %%%%%%%%%%%%%%%%%%%%%%%%%%%%%%%%%%%%%%%%%%%%%%%%%

Since in general the energy of a region with constant external potential
can be written
$ E(S,V,N;\varepsilon) = E(S,V,N) + N \varepsilon $,
and since the number derivative of this is the chemical potential,
a slowly varying one-body potential
can be incorporated into a local chemical potential
(cf. Attard 2025e Eq.~(2.5), de Groot and Mazur 1984).
In the present case for bosonic pairs this is
\begin{equation}
\mu_2({\bf r})
=
\mu_2^{(0)}
- \frac{\beta \mu_{\rm B}^2 \overline{q}_2^2}{3}(\nabla B({\bf r}))^2   ,
\end{equation}
where $\mu_2^{(0)}$ is the chemical potential %for bosonic pairs
in the bare system in the absence of any magnetic field.

The thermodynamic principle that determines superfluid flow
is that energy is minimized at constant entropy
(Attard 2022e, 2025a \S\S~4.3 and 4.6).
At equilibrium this
is equivalent to the local chemical potential being the same
in all connected superfluid regions,
\begin{equation}
\mu({\bf r}) = \mu.
\end{equation}
This gives the fountain pressure equation of H. London (1939),
%for two chambers held at different temperatures,
%$\mathrm{d}p_{\rm BA}/\mathrm{d}T_{\rm B} = \rho_{\rm B} s_{\rm B}$,
%where $p_{\rm BA}$ is the pressure difference, $T$ is the temperature,
%$\rho$ is the number density, and $s$ is the entropy per particle.
which has been confirmed by countless laboratory measurements.
It is manifest in the extraordinary efficiency of superfluidity
in eliminating temperature and chemical potential inhomogeneities.
The difference between ordinary and superfluid equilibrium thermodynamics
is that %for the exchange of particles,
ordinary systems must have equal chemical potential divided by temperature,
whereas superfluid systems must have equal chemical potential;
the former corresponds to maximum total entropy,
whereas the latter corresponds to minimum subsystem energy
at constant entropy.
Also, Bose-Einstein condensation is non-local,
which drives the uniform, incompressible density of the condensed bosons.
The thermodynamic principle for superfluid flow
has recently (Attard 2025e) been used
to derive the well-known two-fluid equations for superfluid hydrodynamics
(Tisza 1938).

As a general thermodynamic principle
%which has been confirmed for superfluid flow,
it must also apply to superconductor currents.
In this case connected regions
must have %the same bosonic pair chemical potential,
$\mu_2({\bf r}) = \mbox{const}.$,
or
\begin{equation}
 \nabla {B}({\bf r}) = \mbox{const}.
\end{equation}
Since in macroscopic volumes the magnetic field would diverge
if it had constant gradient everywhere,
this must be zero.
Hence the magnetic field itself must  be constant,
\begin{equation}
{\bf B}({\bf r})
=
(1+\chi) {\bf B}_{\rm ap}({\bf r})
= \mbox{const}.
%= {\bf B},
\end{equation}
Since this must hold for applied magnetic fields
with arbitrary spatial variation,
and since the  magnetic susceptibility
has to be a property of the superconductor
that is independent of the applied field,
this gives
\begin{equation}
\chi = -1,
\mbox{ and }
{\bf B}({\bf r}) = {\bf 0} .
\end{equation}
This was what was to be proved.
It is the Meissner-Ochsenfeld (1933) effect.

In type I superconductors there is a critical magnetic field
beyond which the applied field fully penetrates the sample, $\chi = 0$,
and superconductivity is destroyed.
This gives the stabilization energy
of the superconducting state (Annett 2004, Kittel 1976, Tinkham 2004).
%Presumably this corresponds to the field
%at which the undiminished applied magnetic force,
%%what force if $B_{\rm ap}$ is uniform?
%%%maybe in the surface region where the field decays?
%which is equal and opposite for each electron in the bosonic pair,
%exceeds the maximum gradient of the potential well that binds them.

In type II superconductors there is a lower
and an upper critical magnetic field,
between which the field partially penetrates the sample
in the form of quantized flux tubes of normal fluid
surrounded by superconducting vortex currents of zero vorticity,
separated by regions of superconductivity with no nett field
(Annett 2004, Kittel 1976, Tinkham 2004).
Presumably in such materials this is more favorable thermodynamically
than the complete exclusion of the field.
%Presumably the upper critical field
%is the point of close packing of the flux tubes.

%%%%%%%%%%%%%%%%%%%%%%%%%%%%%%%%%%%%%%%%%%%%%%
\subsubsection{Modified London Equation}

For superfluid flow,
the rate of change of the momentum density
for condensed bosons is proportional to
the gradient of the chemical potential,
(Attard 2025e Eq.~(2.18))
\begin{equation} \label{Eq:dotp0}
\frac{\partial {\bf p}_0}{\partial t}
=
- n_0 \nabla \mu - \nabla \cdot ({\bf p}_0 {\bf v}_0 ).
\end{equation}
Here and below all quantities
are functions of position ${\bf r}$ and time $t$.
As mentioned, this result gives the two-fluid equations of superfluid flow,
which, amongst other things,
successfully predicted second sound
(Tisza 1939, Landau 1941).

In the present case of superconductivity,
the momentum density for the condensed bosonic pairs
involves the canonical momentum, % (changed from version 1),
which includes the contribution of the magnetic field,
$ {\bf p}_{20} = 2m n_{20} {\bf v}_{20}
- 2 e n_{20} {\bf A}$,
where $n_{20}$ is the number density of the pairs
and ${\bf v}_{20}$ is their local velocity.
The magnetic field is given by the magnetic vector potential,
${\bf B} = \nabla \times {\bf A}$.
The superconducting current is ${\bf j}_{20} = -2 e n_{20}{\bf v}_{20}$.
The chemical potential for the condensed bosonic pairs is
$ \mu_2 = \mu_2^{0}
- {\beta \mu_{\rm B}^2 \overline{q}_2^2} (\nabla B)^2/{3}
-2e \phi$,
where $\phi$ is the electrostatic potential.
This is the electrochemical potential (de Groot and Mazur 1984 Eq.~(XIII.42))
with the  magnetic quadrupole contribution added,
and no velocity-dependent terms.
%(changed from version 1).

With these
Eq.~(\ref{Eq:dotp0}) becomes
\begin{eqnarray}
\lefteqn{
\frac{\partial {\bf j}_{20}}{\partial t}
+ \frac{2e^2}{m}
\frac{\partial (n_{20} {\bf A}) }{\partial t}
} \nonumber \\
& = &
\frac{ -e\beta \mu_{\rm B}^2 \overline{q}_2^2}{3m}
n_{20} \nabla (\nabla B)^2
- \frac{2e^2}{m} n_{20} \nabla \phi .
\end{eqnarray}
Here we have neglected the convective term
$(e/m)\nabla \cdot ({\bf p}_{20} {\bf v}_{20} )$
%= (e/m) \nabla \cdot (  2 m n_{20} {\bf v}_{20} {\bf v}_{20}
%- 2 e  n_{20} {\bf A}  {\bf v}_{20})$
on the grounds that in the linear regime it is small,
although one could argue to keep it.

Taking the curl of this equation,
and using the facts that, by a Maxwell equation,
$\nabla \times {\bf B} = \mu_0 {\bf j}_{20}$,
as well as
$\nabla \times \nabla \times {\bf B} = - \nabla^2 {\bf B}$,
the left hand side is
\begin{eqnarray}
\lefteqn{
\nabla \times
\left[ \frac{\partial  {\bf j}_{20}}{\partial t}
+
\frac{ 2e^2}{m}
\frac{\partial ( n_{20} {\bf A})  }{\partial t}
\right]
}  \\
& =&
% - \mu_0^{-1} \frac{\partial   \nabla^2 {\bf B} }{\partial t}
% + \frac{2e^2 }{m} \frac{\partial( n_{20}{\bf B}) }{\partial t}
% - \frac{2e^2}{m} \frac{\partial ({\bf A}\times \nabla n_{20})}{\partial t}
%\nonumber \\ & =&
\frac{\partial }{\partial t}
\left[ - \mu_0^{-1} \nabla^2 {\bf B}
+
\frac{ 2e^2 n_{20}}{m} {\bf B}
\right]
- \frac{2e^2}{m} \frac{\partial ({\bf A}\times \nabla n_{20})}{\partial t} .
\nonumber
\end{eqnarray}
Since the curl of the gradient of a scalar is zero,
the right hand side is
\begin{eqnarray}
\lefteqn{
\nabla \times
\left[
\frac{-e\beta \mu_{\rm B}^2 \overline{q}_2^2}{3m}
n_{20}\nabla (\nabla B)^2
-\frac{2e^2}{m} n_{20}\nabla \phi
\right]
} \nonumber \\
& =&
\left[ \frac{e\beta \mu_{\rm B}^2 \overline{q}_2^2}{3m}\nabla (\nabla B)^2
+ \frac{2e^2}{m} \nabla \phi
\right] \times \nabla n_{20} .
\end{eqnarray}
Equating these and rearranging  gives
\begin{eqnarray}
\lefteqn{
\frac{\partial }{\partial t}
\left[ - \mu_0^{-1} \nabla^2 {\bf B}
+
\frac{ 2e^2 n_{20}}{m} {\bf B}
\right]
} \nonumber \\
& =&
\left[ \frac{e\beta \mu_{\rm B}^2 \overline{q}_2^2}{3m}\nabla (\nabla B)^2
+ \frac{2e^2}{m} \nabla \phi
\right] \times \nabla n_{20}
\nonumber \\ & & \mbox{ }
+ \frac{2e^2}{m} \frac{\partial ({\bf A}\times \nabla n_{20})}{\partial t} .
\end{eqnarray}
%The right hand side differs from version 1.
This is essentially the same as the first London equation
except that the right hand side
%(which differs from version 1),
is non-zero.
%(Of course the London brothers used the charge and mass of an electron
%and the electron density,
%whereas we use the charge and mass of an electron pair, and the density
%of condensed bosonic pairs.)
The London brothers (1935 Eqs~(2)--(5)),
and apparently everyone ever since
(Annett 2004, Kittel 1976, Tinkham 2004),
assumed that the superconducting electron density
was constant in time and uniform in space.
This assumption gives the constant penetration length often denoted
$\Lambda_{\rm L} = (2 \mu_0 e^2 n_{20}/m )^{-1/2}$.
But this assumption is not reasonable during the superconducting transition.
If a sample in a magnetic field
is cooled below  the superconducting transition temperature,
then obviously the density of superconducting electrons must go from zero
to some finite value, which is to say that it is time dependent.
Also, in the process of the transition
the superconducting electrons are nucleated at different points in space
due to temperature and magnetic field inhomogeneities,
which means that $\nabla n_{20}({\bf r},t) \ne {\bf 0}$.
Indeed the magnetic quadrupole energy for bosonic electron pairs
is one source of nucleation inhomogeneity.
%Finally, a type II superconductor with partial magnetic field penetration
%has inhomogeneous superconducting electron density.

The first London equation,
which is this equation with right hand side zero,
has always been rejected on the grounds
that it would trap a pre-existing magnetic field in place
after the superconducting transition
(Annett 2004, Kittel 1976, F and H London 1935, Tinkham 2004),
which is contrary to the Meissner-Ochsenfeld (1933) effect.
That conclusion assumes
uniformity in space and time for the superconducting electron density.
And it invokes the Drude equation with zero resistivity,
in contrast to the superfluid equation (\ref{Eq:dotp0}).
The present argument casts doubt on the physical basis for both assumptions.
The present result shows that if the density of condensed bosonic pairs
is non-uniform during the transition,
then the magnetic field can change with time
and need not be trapped within the superconductor.

Of course after equilibration the magnetic field is independent of time
and both sides of this equation must be zero.
In this case the gradient of the magnetic field
and the gradient of the potential vanish
in the macroscopic interior of the sample.
In the surface region
the gradient of the condensed bosonic pair density either vanishes
or else lies parallel to the other gradients.
%(Because condensation is non-local,
%the present author expects it to vanish (Attard 2025a \S5.4.1).)

Setting the right hand side to zero, and integrating over time gives
\begin{equation} %\label{Eq:dotp0}
%\frac{\partial }{\partial t}\left[
 -\mu_0^{-1}  \nabla^2 {\bf B}({\bf r})
+
\frac{ 2e^2 n_{20}}{m} {\bf B}({\bf r})
%\right]
=
{\bf C}({\bf r}).
%\frac{e n_{20} \mu_{\rm B} \overline{q}_2}{m} \nabla^2 | \nabla {B}|.
\end{equation}
But from the results in the previous section,
${\bf C}({\bf r}) \to 0$ in the interior of the superconductor.
Choosing the initial condition ${\bf C}({\bf r}) = {\bf 0}$ everywhere
gives a particular solution that is consistent with the
Meissner-Ochsenfeld (1933) effect.
This is the second London equation,
and it is believed to give
the decay of the magnetic field in the surface region
(Annett 2004, Kittel 1976, F and H London 1935, Tinkham 2004).

% end comment %%%%%%%%%%%%%%%%%%%%%%%%%%%%%%%%%%%%%%%%%%%%%%%%

%%%%%%%%%%%%%%%%%%%%%%%%%%%%%%
\subsubsection{Critical Magnetic Field}

This section identifies the physical mechanism
by which superconductivity is destroyed
when the magnetic field exceeds a critical value.
We use an ideal electron model,
which, in the linear regime, can be shown to give
the known result for the Pauli paramagnetic susceptibility
(Pathria 1972 \S8.2).
We combine this with a treatment of Bose-Einstein condensation
using ideal bosons,
similar to that used by F. London (1935,
Attard 2025a Ch.~2,  Pathria 1972 \S7.1).

The unpaired electrons, labeled $1\pm$,
are neglected  for the present purposes.
The paired electrons consist of fermionic pairs
in which both electrons have the same spin, $2\pm$,
and bosonic pairs in which the electrons have opposite spin,
and which are either condensed, $20$, or uncondensed, $2*$.
There is no prohibition on electrons in a fermionic pair
having the same spin as they are in different momentum states.
Also the Fermi repulsion
(cf.\ Attard 2025a \S2.2.5, Pathria 1976 \S5.5)
does not apply because the momentum is not integrated over.
The propensity to form electron pairs is determined
by the characteristics of the material and the thermodynamic state,
and is different for low- and for high-temperature superconductors
(Annett 2004, Attard 2025a \S6.5,
Bardeen \emph{et al.}\ 1957,  Tinkham 2004).

We treat the case that the applied magnetic field
partially or entirely penetrates the sample,
${\bf B} = [1+\chi] {\bf B}_{\rm ap}$,
with $\chi > -1 $.
Since the goal is restricted to discovering the electronic mechanism
by which superconductivity is destroyed,
for simplicity we take the magnetic field
to be uniform over the region being considered.
This means that we can neglect the magnetic quadrupole contribution
of the paired electrons.
We do not consider the effects of magnetic field heterogeneity,
quantized flux tubes,
%vortex-free superconductor current vortices,
etc.
The effective fugacity for the unpaired electrons is
$z_\pm = z e^{\pm \beta \mu_0 \mu_{\rm B} B}$,
and that for the paired electrons is
$z_{2\pm} =  z_2 e^{\pm 2\beta \mu_0 \mu_{\rm B} B}$.
Below the superconducting transition,
$z_2 \equiv e^{\beta \mu_2^{(0)}} = 1^-$.
It is emphasized that this is an artefact of the ideal boson model.

By standard methods
(cf.\ Pathria 1972 \S8.2),
the average number of fermionic electron pairs is
\begin{equation}
\overline N_{2\pm}(z_{2\pm},V,T)
%=
%\frac{z_{2\pm} \partial (-\beta \Omega_{2\pm})}{\partial z_{2\pm}  }
=
V 2^{-3/2} \Lambda^{-3} f_{3/2}(z_{2\pm}) ,
\end{equation}
where the Fermi-Dirac integral appears (Pathria 1976 Appendix~E).
The thermal wavelength for single electrons is
$\Lambda = \sqrt{2\pi \beta \hbar^2 /m}$.
Similarly
(cf.\ Attard 2025a Ch.~2, Pathria 1972 \S7.1),
the average number of uncondensed bosonic electron pairs is
\begin{eqnarray}
\overline N_{2*}(z_2,V,T)
& = &
V 2^{-3/2} \Lambda^{-3}   g_{3/2}(z_{2}),
%\nonumber \\ & \to &
%V 2^{-3/2} \Lambda^{-3}  g_{3/2}(1). % \zeta({3/2}),
\end{eqnarray}
where the Boise-Einstein integral appears (Pathria 1976 Appendix~D).
It is an artefact of the ideal boson model
that at a given temperature this has a maximum value,
$g_{3/2}(1) = \zeta(3/2) =2.612$.
This average number is independent of the magnetic field
and below the transition
it is insensitive to the actual value of the pair fugacity.

The number of condensed bosonic electron pairs is
\begin{equation}
\overline N_{20} %(z_2,V,T)
= N_2 - \overline N_{2+} - \overline N_{2-} -  \overline N_{2*} .
\end{equation}
This is used below the superconducting transition.
Given the fixed number of electrons pairs $N_2$,
from measurement or other,
this determines the pair fugacity,
$z_2 = \overline N_{20}/[1+\overline N_{20}] \to 1^-$.
This is an artefact of the ideal electron model
and the treatment is analogous to that of F. London (1938)
for superfluidity (Attard 2025a Ch.~2, Pathria 1972 \S7.1).
We could include the unpaired electrons in this
without changing the conclusion.

At high fields the grand potential is dominated by $\Omega_{1+}$
and by $\Omega_{2+}$,
which favor the penetration of the magnetic field.
One can show that the total number of fermionic electron pairs
increases with increasing magnetic field,
\begin{equation}
\frac{\partial [\overline N_{2+} + \overline N_{2-}] }{\partial B}
> 0,
\end{equation}
and similarly for the unpaired electrons.
This is a non-linear effect
in which the increase in spin-up electrons is greater than
the decrease in spin-down electrons.
Since the  relatively small number of uncondensed bosonic pairs,
$\overline N_{2*} $,
is independent of the magnetic field,
and since the total number of pairs is determined
by the material and the thermodynamic state,
this shows that the number of condensed bosonic electron pairs
must decrease with increasing magnetic field,
${\partial \overline N_{20}  }/{\partial B} < 0$.
Hence there exists a critical field
at which the number of condensed bosonic electron pairs goes to zero
and
%Since  the superconducting current is carried by
%condensed bosonic electron pairs,
superconductivity is annihilated.

%Including unpaired electrons is not expected
%to change qualitatively this result.
%Since we have assumed a uniform magnetic field,
%the magnetic quadrupole contribution
%of the bosonic pairs does not contribute.

The conclusion is that a magnetic field destroys superconductivity
because it non-linearly favors spin-up electrons,
which reduces the number of spin-down electrons available
for bosonic electron pairs.
When the bosonic electron pair density  falls below
their transition density in the absence of a field,
then superconductivity is destroyed.

%\newpage $\;$ \newpage
%%%%%%%%%%%%%%%%%%%%%%%%%%%%%%%%%%%%%%%%%%%%%%%%%%%%%%%%%%%%%%%%%%%%%%%%%%
\vspace{1cm}
\section*{References}
%\vspace{-1cm}

%%%%%%%%%%%%%%%%%%%%%%%%%%%%%%%%%%%%%%%%%%%%%%%%%%%%%%%%%%%%%%%%%%%%%%%%%%

\begin{list}{}{\itemindent=-0.5cm \parsep=.5mm \itemsep=.2mm} % .5mm

\item % refd
Annett J E 2004
\emph{Superconductivity, Superfluids and Condensates}
(Oxford: Oxford University Press)

\item  % Attard22b % QSM22/paper1 % refd
Attard P 2022b
Attraction between electron pairs in high temperature superconductors
arXiv:2203.02598

\item  % refd
Attard P 2022e
Further on the fountain effect in superfluid helium
arXiv:2210.06666

\item % refd
Attard P 2025a
\emph{Understanding Bose-Einstein Condensation,
Superfluidity, and High Temperature Superconductivity}
(London: CRC Press)

%\item % Attard25b % QSM24/SFVisco %Appendix
%Attard P 2025b
%The molecular nature of superfluidity: Viscosity of helium from quantum
%  stochastic molecular dynamics simulations over real trajectories
%arXiv:2409.19036v5

%\item % Attard25d % QSM25/lambda %Appendix
%Attard P 2025d
%Bose-Einstein condensation and the lambda transition
%for interacting Lennard-Jones helium-4
%arXiv:2504.07147v1

\item % Attard25e % QSM25/TwoFluid % refd
Attard P 2025e
The two-fluid theory for superfluid hydrodynamics
and rotational motion
arXiv:2505.08826v5

\item %refd
Bardeen J, Cooper L N, and Schrieffer J R 1957
Theory of superconductivity
\emph{Phys.\ Rev.}\ {\bf 108} 1175

\item %refd
Cooper L  N 1956
Bound electron pairs in a degenerate Fermi gas
\emph{Phys.\ Rev.}\ {\bf 104} 1189

\item %refd
de Groot S R and Mazur P 1984
\emph{Non-equilibrium Thermodynamics}
(New York: Dover)

\item  % refd
Kittel C 1976
\emph{Introduction to solid state physics}
(New York: Wiley 5th edn)

\item % refd
Landau L D 1941
Theory of the superfluidity of helium II
\emph{Phys.\ Rev.}\ {\bf 60} 356.
Two-fluid model of liquid helium II
\emph{ J.\ Phys.\ USSR} {\bf 5} 71

\item %refd
London F and London H 1935
The electromagnetic equations of the supraconductor
\emph{Proc.\ Royal Soc.\ (London)} {\bf A149} 72.
Supraleitung und diamagnetismus
\emph{Physica} {\bf 2} 341. %-354

\item  % refd
London F 1938
The $\lambda$-phenomenon of liquid helium and the Bose-Einstein degeneracy
\emph{Nature} {\bf 141} 643

\item %refd
London H 1939
Thermodynamics of the thermomechanical effect of liquid He II
\emph{Proc.\ Roy.\ Soc.}\ {\bf  A171} 484

\item % refd
Meissner W and Ochsenfeld R 1933
Ein neuer effekt bei eintritt der supraleitf\"ahigkeit
\emph{Die Naturwissenschaften} {\bf 21} 787

\item % refd
Pathria R K 1972
\emph{Statistical Mechanics} (Oxford: Pergamon Press)

\item % refd
Tinkham M 2004
\emph{Introduction to Superconductivity}
(New York: Dover 2nd edn)

\item  %refd
Tisza L 1938
Transport phenomena in helium II
\emph{Nature} {\bf 141} 913

%%%%%%%%%%%%%%%%%%%%%%%%%%%%%%%%%%%%%%%%%%%%%%%

%\item
%\ldots

%\item % refd
%Attard  P 2012
%\emph{Non-equilibrium Thermodynamics and Statistical Mechanics:
%Foundations and Applications}
%(Oxford: Oxford University Press)

\end{list}

%%%%%%%%%%%%%%%%%%%%%%%%%%%%%%%%%%%%%%%%%%%%%%%%%%%%%%%

%\newpage
%%%%%%%%%%%%%%%%%%%%%%%%%%%%%%%%%%%%%%%%%%%%%%%%%%%%%%%%%%%%%%%%%%%%%

%\appendix
%\newpage
%%%%%%%%%%%%%%%%%%%%%%%%%%%%%%%%%%%%%%%%%%%%%%%%%%%%%%%%%%%%%%%%%%%%%%%%%%
%
%\section{Critical Field} %\label{Sec:ds/de-etc}
%\setcounter{equation}{0} \setcounter{subsubsection}{0}
%%\renewcommand{\theequation}{\arabic{section}.\arabic{equation}}
%\renewcommand{\theequation}{\Alph{section}.\arabic{equation}}
%
%%%%%%%%%%%%%%%%%%%%%%%%%%%%%%%%%%%%%%%%%%%%%%%%%%%%%%%%%%%%%%%%%%%%%%%%%%

\end{document}